\documentclass[a4paper,
              ]{jpconf}
 {\usepackage[utf8]{inputenc}}           % switch to utf8		 
\usepackage[british,english,american]{babel}
\usepackage[final]{pdfpages}
\usepackage{multirow}
\usepackage{ragged2e}
\usepackage[caption=false]{subfig}
\graphicspath{{figures/}}

\usepackage{setspace}
\begin{document}
\doublespacing
\title{High-quality positrons from a multi-proton bunch driven hollow plasma wakefield accelerator}
%\thanks{Work supported by the PDS Award of The University of Manchester, the Cockcroft Institute core grant, STFC, and SB RAS grant 0305-2017-0021.}}

\author{Y. Li\textsuperscript{1,2}, G. Xia\textsuperscript{1,2}, K. V. Lotov\textsuperscript{3,4}, A. P. Sosedkin\textsuperscript{3,4}, Y. Zhao\textsuperscript{1}}
\address{\textsuperscript{1}University of Manchester, Manchester, UK \\ 
\textsuperscript{2}Cockcroft Institute, Daresbury, UK\\
\textsuperscript{3}Budker Institute of Nuclear Physics, Novosibirsk, Russia\\
\textsuperscript{4}Novosibirsk State University, Novosibirsk, Russia}
\ead{yangmei.li@manchester.ac.uk}	
\vspace{10pt}
\begin{indented}
\item[]September 2018
\end{indented}

\begin{abstract}
By means of hollow plasma, multiple proton bunches work well in driving nonlinear plasma wakefields and accelerate electrons to energy frontier with preserved beam quality. However, the acceleration of positrons is different because the accelerating structure is strongly charge dependent. There is a discrepancy between keeping a small normalized emittance and a small energy spread. This results from the conflict that the plasma electrons used to provide focusing to the multiple proton bunches dilute the positron bunch. By loading an extra electron bunch to repel the plasma electrons and meanwhile reducing the plasma density slightly to shift the accelerating phase with a conducive slope to the positron bunch, the positron bunch can be accelerate to 400\,GeV (40$\%$ of the driver energy) with an energy spread as low as 1$\%$ and well preserved normalized emittance. The successful generation of high quality and high energy positrons paves the way to the future energy frontier lepton colliders. 
\end{abstract}

\begin{indented}
\item[]\textbf{Keywords:} high quality, positron acceleration, multiple proton bunches, hollow plasma, particle-in-cell
\end{indented}

\section{Introduction}
High energy particle colliders are the main research tools for physicists to study new particles and explore the fundamental structures of the physical world. As leptons are point-like, fundamental objects, a significantly cleaner collision environment is achievable in lepton colliders than in hardon colliders and hence higher precision of physics measurements. As a result, it is widely accepted that the next energy frontier colliders should collide electrons and positrons in the TeV scale. Given the fact that the achievable acceleration gradient in conventional radio-frequency accelerators is subject to electrical breakdown, plasma wakefield accelerators were proposed \cite{tajima1979laser} thanks to their orders of magnitude higher field threshold, and have become increasingly favored in constructing compact and cost-effective colliders in the energy level of hundreds of GeV or TeV \cite{lee2002energy,adli2016proton,hogan2016electron}.      

There has been essential progress in plasma-based electron acceleration to high energy \cite{blumenfeld2007energy} or with high efficiency \cite{litos2014high}, but little advance for positron acceleration. The most fundamental obstacle is lack of stable acceleration regime for positrons. More specifically, in the blowout regime of plasma wakefield acceleration (PWFA), the plasma electrons are displaced off axis by the electron driver and then concentrate in the bubble sheath. The entirely rear half of the bubble which is devoid of plasma electrons forms an ideal accelerating structure for the electron beam, \textit{i.e.}, is both uniformly accelerating and linearly focusing along the beam radius \cite{rosenzweig1991acceleration,lu2006nonlinear}. However, the focusing region for positrons is substantially small, which is at the back of the bubble where the plasma electrons collapse on axis and the plasma density is nonuniform. Therefore, the acceleration and focusing for positrons vary radially, which incurs the growth of energy spread and the emittance. Also it requires precise control of the location of positrons \cite{lotov2007acceleration}. In consequence, the idea in ref. \cite{lotov2007acceleration} of accelerating positrons in the nonlinear wake of the electron beam is apparently challenging \cite{wang2009optimization}. Nevertheless, an encouraging experiment in 2015 \cite{corde2015multi} validated that it is feasible to accelerate the tail positrons by the front of the positron beam. By sufficiently self-loading the wake, the rear of the beam sees focusing from the ``sucked-in'' plasma electrons and a flat accelerating field. Despite a low energy spread, the preservation of the beam emittance is still an issue \cite{hogan2003ultrarelativistic,muggli2008halo}.  

Hollow channel plasma was initially proposed to confine the lasers \cite{katsouleas1992laser}. Later it was recognized to strongly benefit the acceleration and quality of positrons \cite{gessner2016demonstration,yi2014positron}. Take the work in ref. \cite{gessner2016demonstration} for example, the hollow plasma dispels the defocusing from the background ions and also nonlinear focusing from the plasma electrons attracted towards the axis. In addition, the longitudinal wakefields become uniform transversely. Here the hollow channel acts as a waveguide. It requires no plasma electrons streaming into the channel and hence no intense drivers, which are supposed to be used for large wakefield excitation. Ref. \cite{yi2014positron} elucidates strongly nonlinear beam-plasma interaction, whereas the ultra-short and dense proton driver is hardly obtainable, because the real proton bunch in practice is tens of cm long. 

Owning to huge energy contents, protons have the potential to accelerate particles to energy frontier over a single plasma stage \cite{li2017high}. In our previous work \cite{li2017multi}, we exploit multiple equidistant proton bunches, which are hypothetically obtained from longitudinal modulation of the long bunch. This eases the challenge in compressing it to one single short proton bunch. Furthermore, the introduction of the hollow channel removes the ion defocusing to the proton bunches. It also overcomes deterioration of the witness beam quality in uniform plasma, which results from the asymmetric plasma response to the positively charged drivers compared to their counterparts. As a consequence, the multiple proton bunches work well in the nonlinear regime sourcing strong plasma fields and accelerating electrons to energy frontier with preserved beam quality. 

In this paper, we investigate the positron acceleration for the first time in the hollow plasma, where multiple proton bunches play as the driver. However, unlike the witness electrons, the positrons see a different and less favorable accelerating structure, as which is strongly charge dependent. The consequence is a discrepancy in terms of the beam quality, that is, either a preserved normalized emittance with a high energy spread or the opposite way. This issue will be explained in Section 2, where afterwards we propose two intuitive approaches which are capable of mitigating the beam quality deterioration to some extent yet not sufficiently. Fortunately, a combined solution is validated to be able to guarantee both high energy gain and high beam quality.  In Section 3, we analyse the dependence of acceleration performance and beam quality on the beam and plasma parameters, followed by the conclusions in Section 4. 

\section{The obstacle to preservation of positron beam quality and solutions}

\subsection{The dilemma in preserving the beam quality}
By means of hollow plasma, multiple proton bunches have been demonstrated in simulations to successfully work in the nonlinear regime and accelerate electrons to $\sim$TeV with preserved beam quality \cite{li2017multi}. In this case, the focusing to proton bunches mainly comes from the plasma electrons attracted into the channel. Likewise, in this work we adopt multiple proton bunches as the driver along with the indispensable part---hollow plasma channel for the nonlinear wake excitation. Quadrupole magnets are used to prevent the head of the first proton bunch from emittance-driven erosion owing to no or weak plasma focusing therein. The proton beam dynamics and the wake characteristics are similar to those in ref. \cite{li2017multi} thus not repeated here. Yet it is worth mentioning that the proton driver and plasma channel have been carefully optimized in order to radially enlarge the acceleration region for positrons which is free from plasma electrons. This is essential as will be seen later that the positrons are severely liable to the interference of plasma electrons penetrating near axis. Given this it also requires the initial radius and emittance of the positron bunch to be small. The positron bunch is initialized with the equilibrium radius, which is calculated based on ref. \cite{lotov2010simulation} assuming positrons are only focused by the external quadrupole fields. Table \ref{table1}  displays the simulation parameters for the beams and plasma. The quasi-static particle-in-cell code LCODE \cite{lotov2003fine,sosedkin2016lcode} based on the 2D cylindrical geometry has been employed due to its high computing efficiency to conduct all the simulations, because the cases dealt with are concerned with long distances and also the simulation grid must be fine enough to resolve the positron beam. 

\begin{table}[!t]
   \caption{Beam and plasma parameters in simulations}
   \begin{center}
   \begin{tabular}{lll}
\br
       \textbf{Parameters} & \textbf{Values}                      &\textbf{Units} \\
\mr
	   \textbf{Initial proton driver:}\\
           Single bunch population                                  & 1.28~$\times$~$10^{10}$       &                \\ %[3pt]
           Energy                         			                 & 1                                               &TeV          \\ 
           Energy spread               		                          & 10$\%$                                     &                 \\ %[3pt]
           Single bunch length                                         & 66                                             &$\mu$m     \\
           Single bunch radius                                         & 68                                              &$\mu$m     \\
           Bunch train period						 & 660                                            &$\mu$m      
           \vspace{4pt} \\
           \textbf{Initial witness positron bunch:}\\
           Population, $N_{\mathrm p}$                              & 1.0~$\times$~$10^{9}$       &                \\ %[3pt]
           Energy, $W_{\mathrm 0}$                        		  & 10                                               &GeV          \\ 
           Energy spread, $\delta W/W$              		  & 1$\%$                                     &                 \\ %[3pt]
           RMS length, $\sigma_z$                          	  & 5                                             &$\mu$m     \\
           RMS radius, $\sigma_r$                          		  & 4                                              &$\mu$m     \\
           Normalized emittance, $\epsilon_n$                  & 0.25                                           &$\mu$m    
            \vspace{4pt} \\
           \textbf{Hollow plasma channel:}\\
           Plasma density, $n_{\mathrm 0}$                         & 5~$\times$~$10^{15}$                &cm$^{-3}$  \\
           Channel radius, $r_{\mathrm c}$                          & 190                                          &$\mu$m  
            \vspace{4pt} \\
           \textbf{Quadrupole magnets:}\\
           Magnetic field gradient, $S$                                 & 500                                          &T/m  \\
          Quadrupole period, $L$                                         & 0.9                                       &m \\
\br
   \end{tabular}
   \end{center}
   \label{table1}
\end{table}

\begin{figure*}[!t]
 \begin{center}
  \subfloat{
   \centering
   \includegraphics[width=38pc]{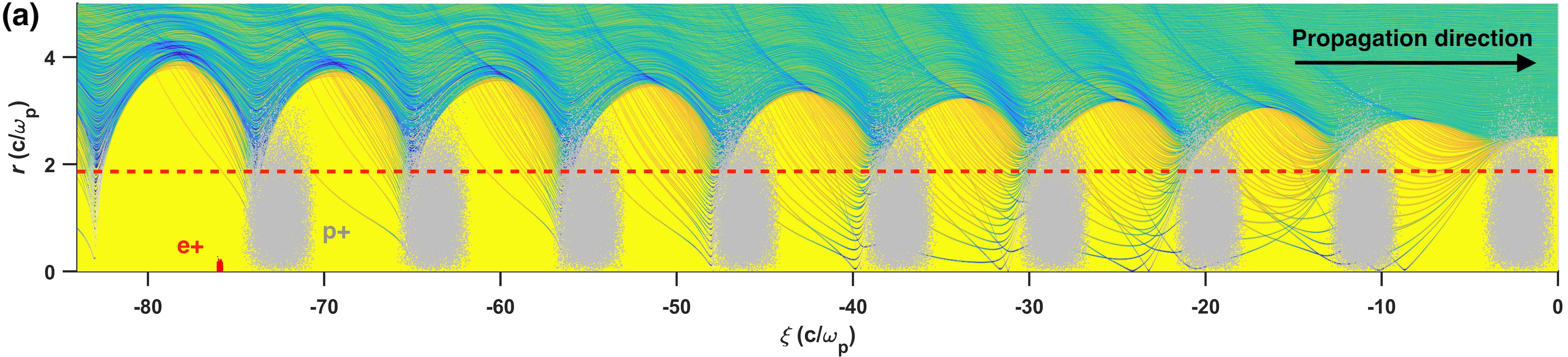}
  }
  \newline
    \subfloat{
   \centering
   \includegraphics[width=40pc]{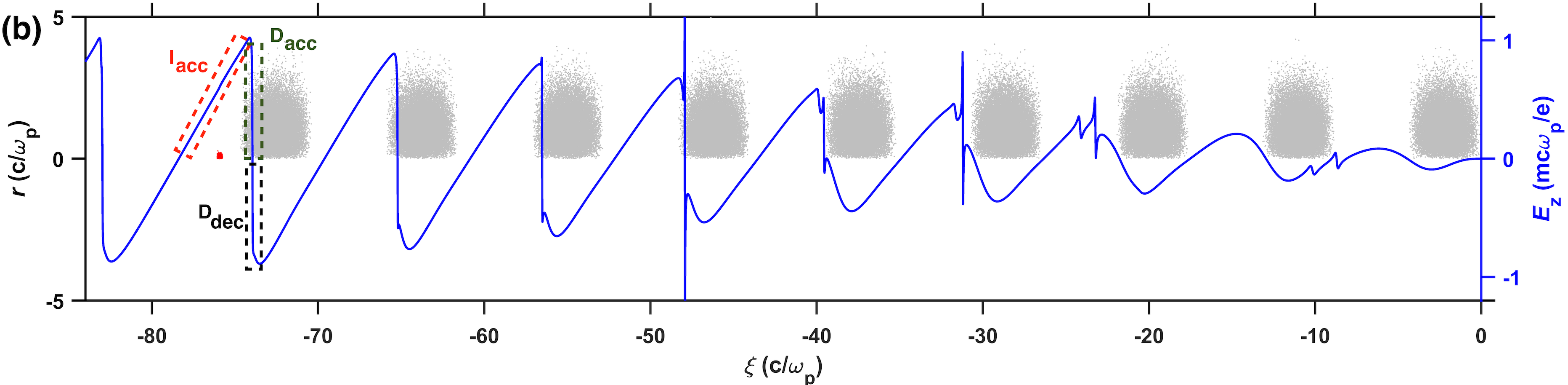}
  }
  \newline
  \subfloat{
   \centering
   \includegraphics[width=9pc]{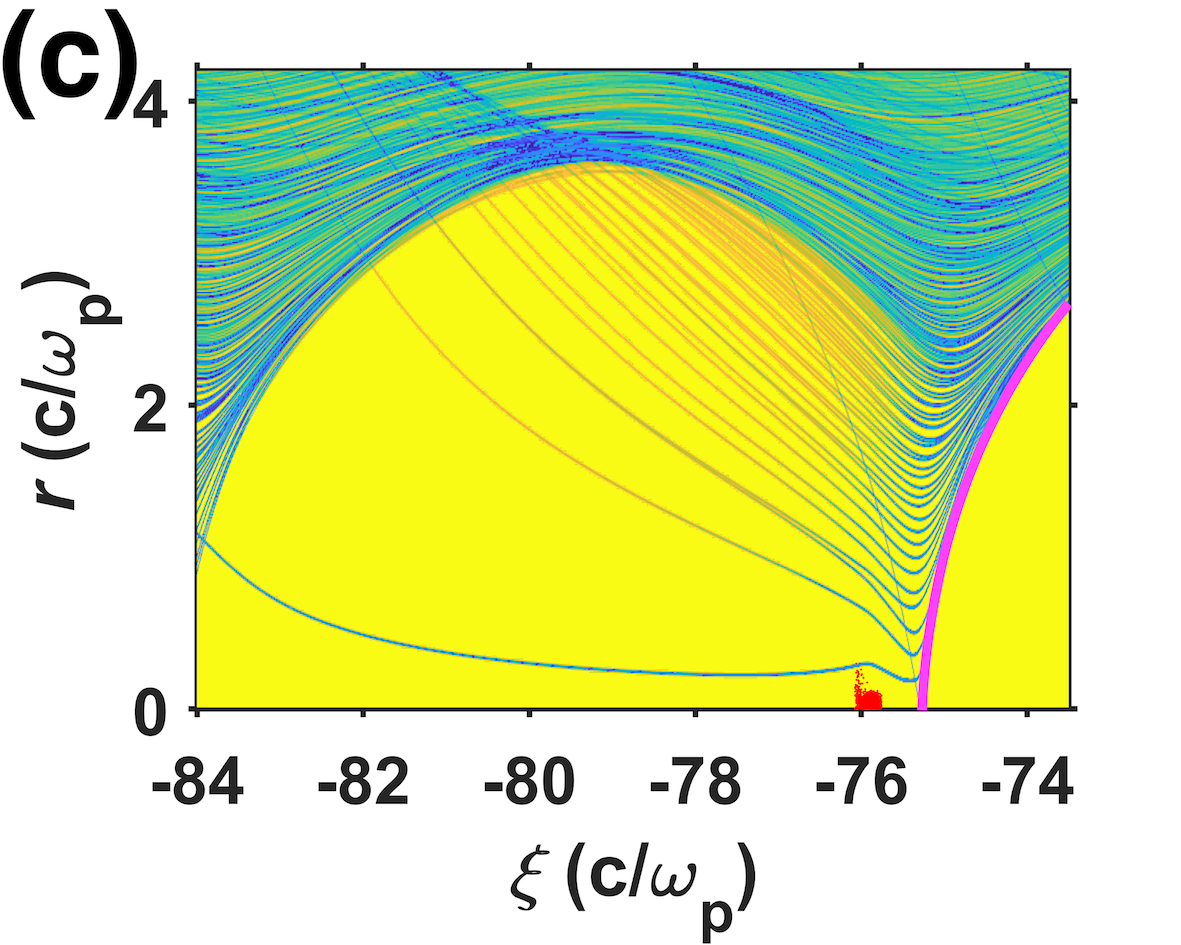}
  }
  \subfloat{
   \centering
   \includegraphics[width=9pc]{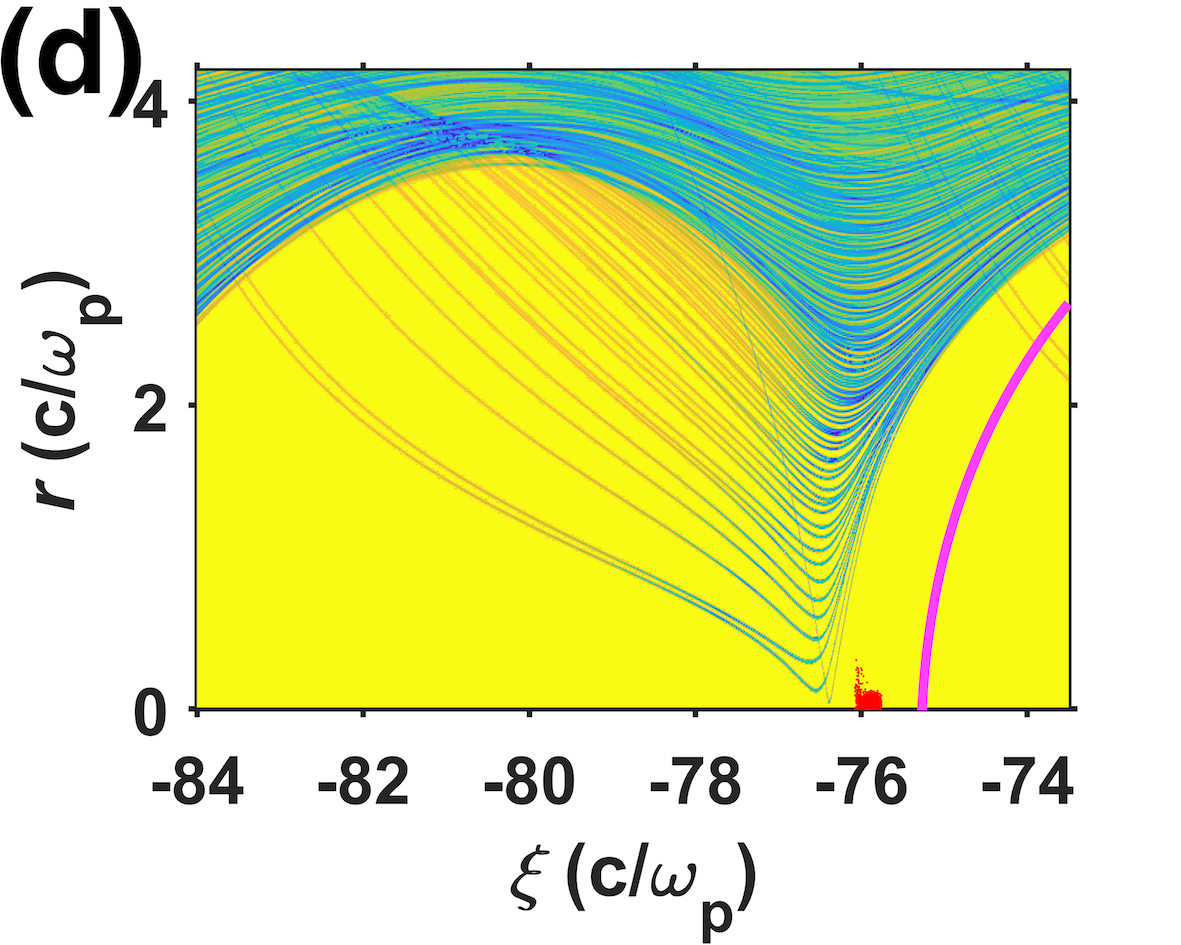}
  }
    \subfloat{
   \centering
   \includegraphics[width=9pc]{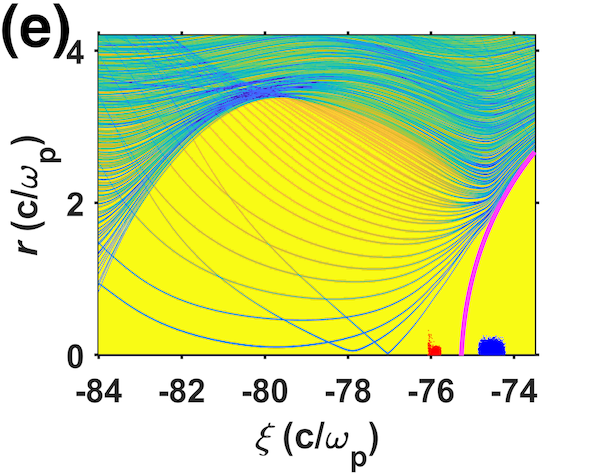}
  }
  \subfloat{
   \centering
   \includegraphics[width=9pc]{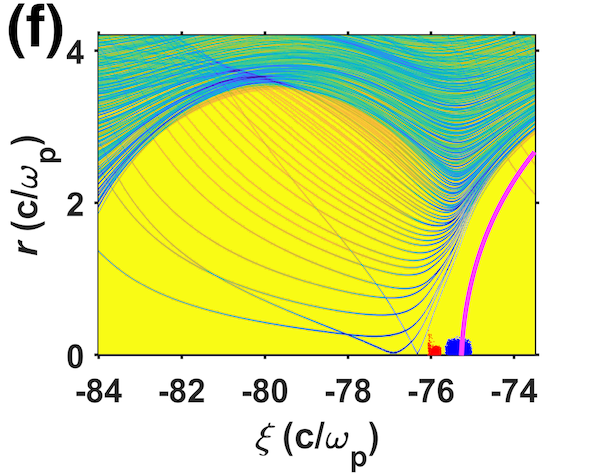}
  }
    \newline
  \subfloat{
   \centering
   \includegraphics[width=9pc]{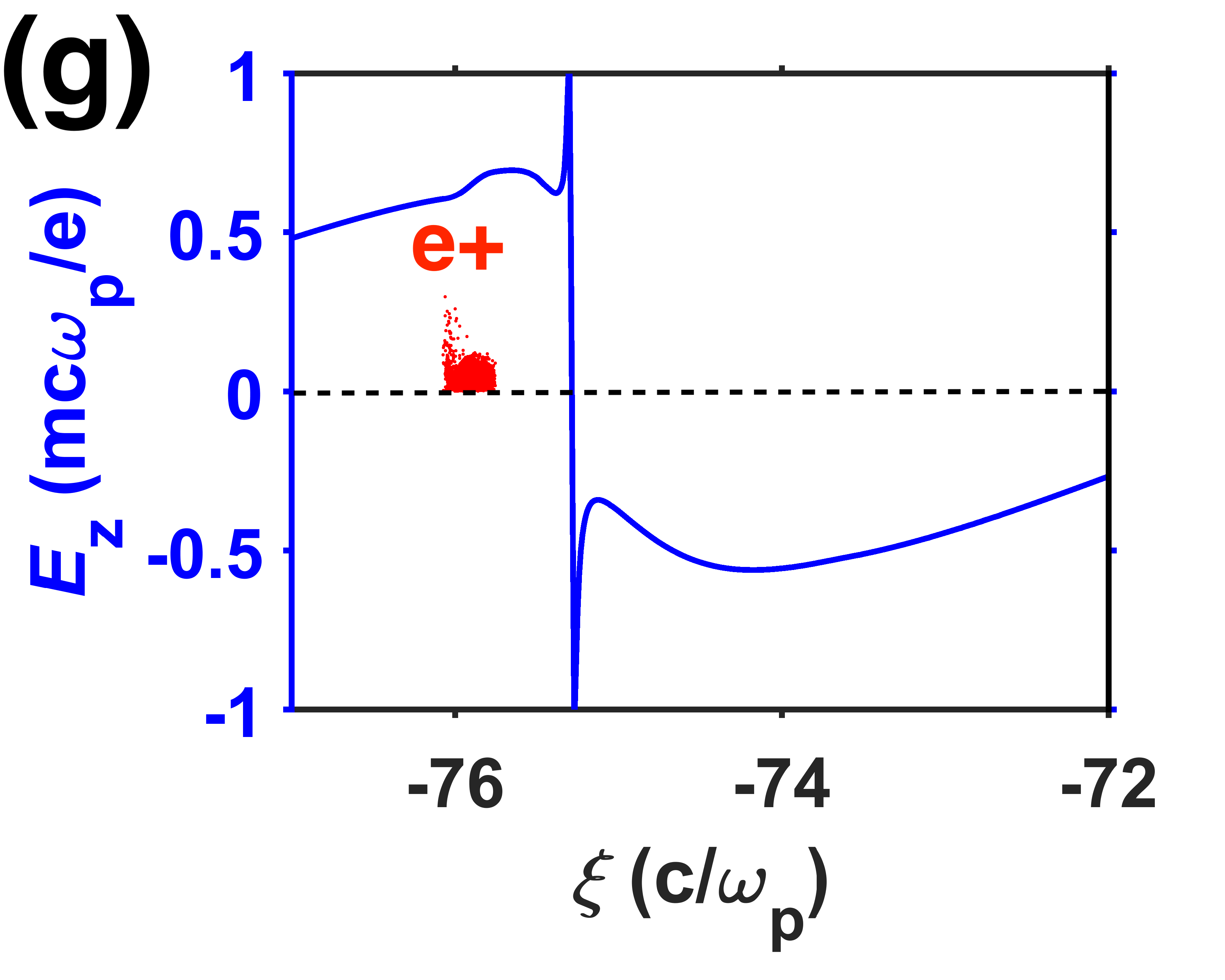}
  }
  \subfloat{
   \centering
   \includegraphics[width=9pc]{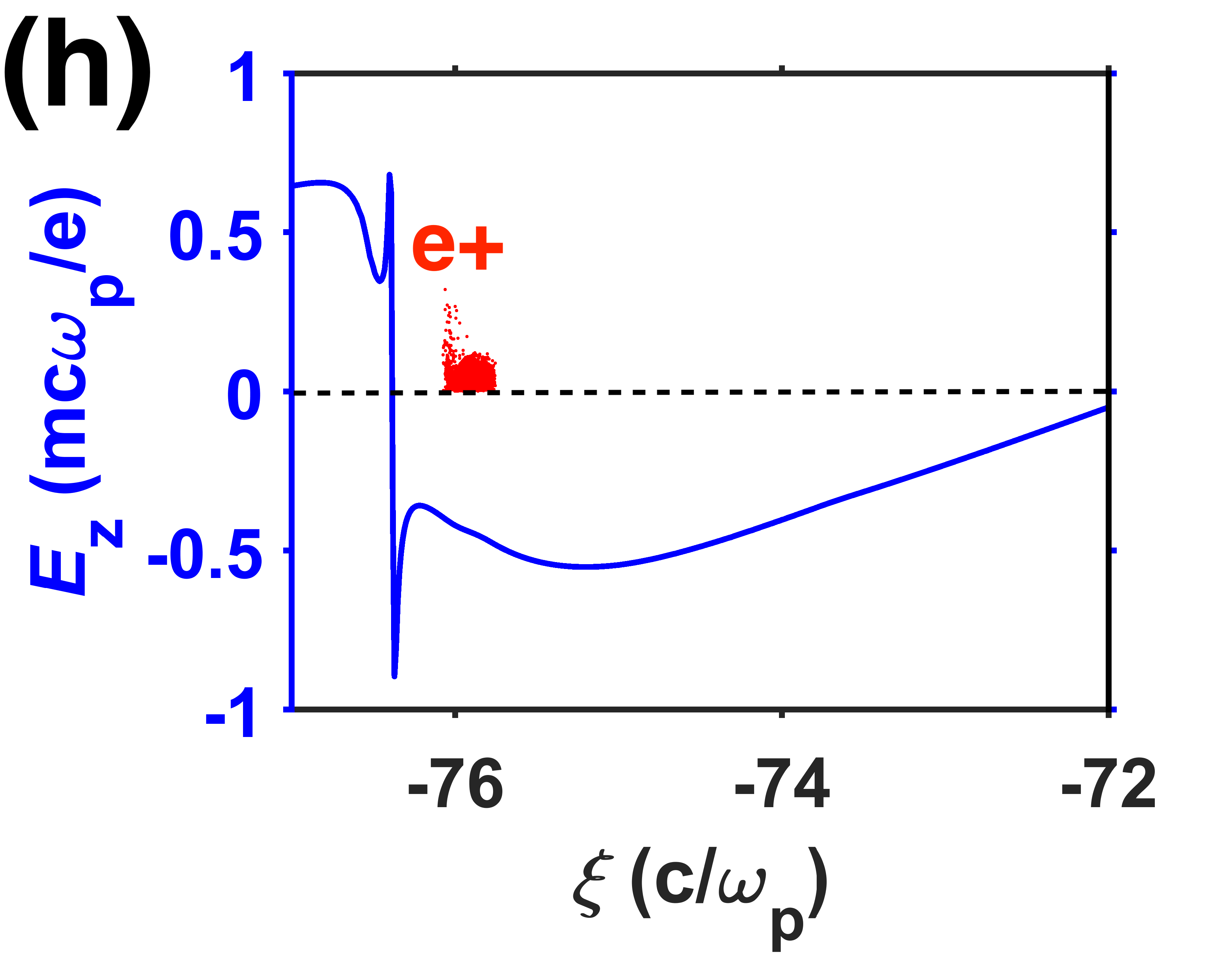}
  }
    \subfloat{
   \centering
   \includegraphics[width=9pc]{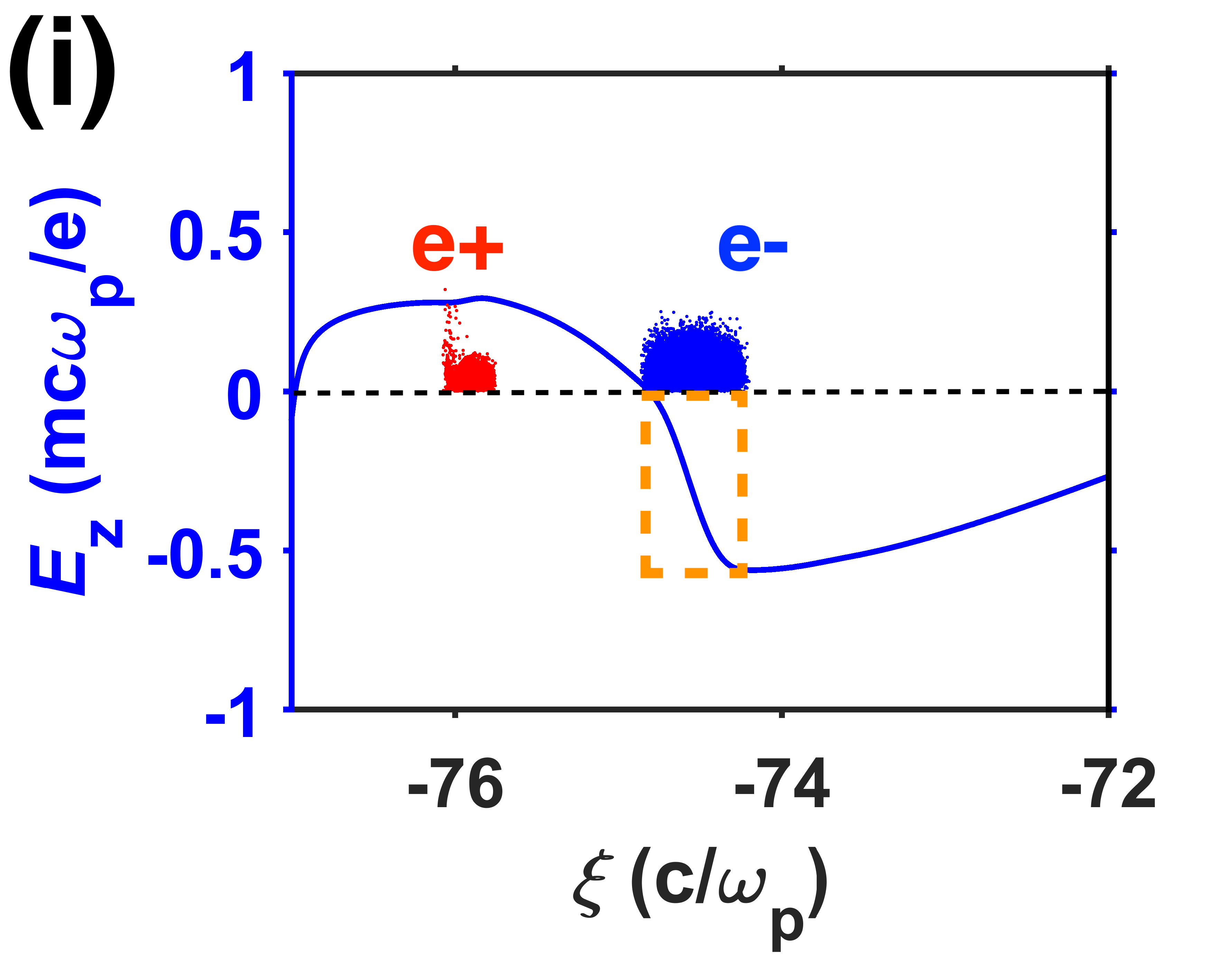}
  }
  \subfloat{
   \centering
   \includegraphics[width=9pc]{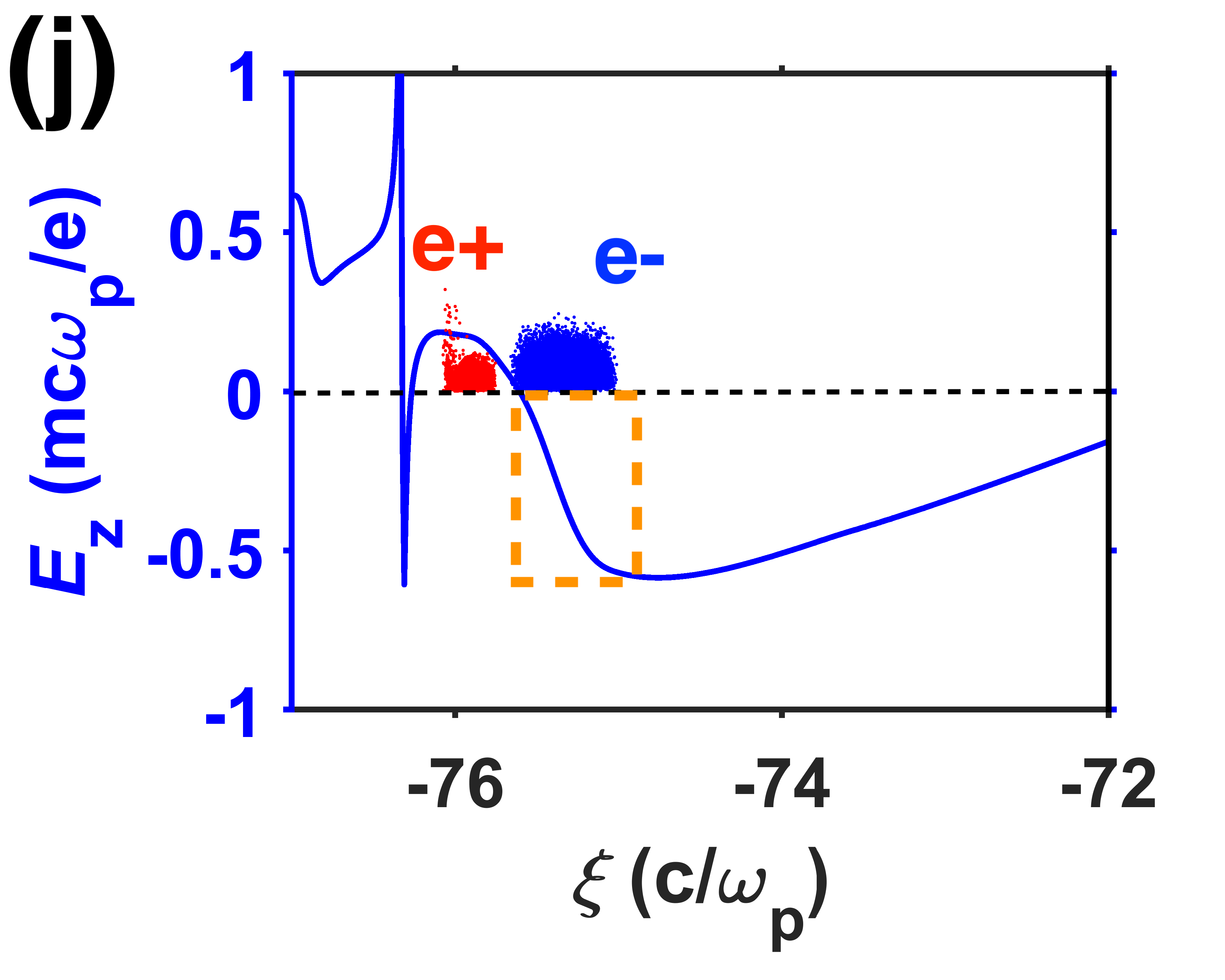}
  }
 \end{center}
 \caption{Axisymmetric space distribution of the plasma electrons, proton and positron bunches (a) and the corresponding on-axis longitudinal electric field (b) at $z=0\,$m, the close-ups of plasma electron distribution (the 3rd row) and the on-axis longitudinal electric field (the 4th row) in the vicinity of the positron bunch at $z=84\,$m for the cases of the original configuration (c, g), decreased by 9$\%$ plasma density (d, h), loading of extra electrons (e, i), and decreased plasma density with electron presence (f, j), respectively. The grey points denote the protons while the red and blue points represent the positrons and extra electrons respectively in all sub-figures. The dashed red line in (a) outlines the hollow channel boundary. The red, green and black frames in (b) mark the ``$\mathrm{I_{acc}}$", ``$\mathrm{D_{acc}}$" and `$\mathrm{D_{dec}}$" regions for positrons, separately. The magneta curves in the third row sketch the boundary of the bubble ahead of the positron bunch under the original configuration. The orange frames in (i) and (j) mark the ``$\mathrm{A_{acc}}$" region for electrons.} 
  \label{fig:ne_Ez_rz}
\end{figure*}

Figure\,\ref{fig:ne_Ez_rz}a shows the axisymmetric space distribution of the plasma electrons (2D map) and proton (grey points) and positron (red points) bunches in a co-moving window where the position coordinate $\xi=z-ct$ is used for convenience. To facilitate comparisons, $\xi$ is normalized to the plasma skin depth $c/\omega_{\mathrm p}$, where $c$ is the light velocity and $\omega_{\mathrm p}$ is the plasma electron frequency corresponding to the initial plasma density of $n_{\mathrm 0}$. There are no plasma ions within the hollow channel. Each proton bunch basically stays in the rear half of the bubble seeing decelerating fields (Fig.\,\ref{fig:ne_Ez_rz}b). The positron bunch is located within the first half of the bubble just behind the last proton bunch (\textit{i.e.}, the 9th one), where the longitudinal field is accelerating and radially uniform. In the initial acceleration stage (before $z=84\,$m), there are no plasma electrons penetrating into the positron region, which makes the positrons only focused by the quadrupole fields. With its initial radius matching with the focusing structure, the positron bunch is stably accelerated with a well-preserved normalized emittance.  
 
Unfortunately, the slope of the accelerating field initially seen by positrons is positive and is not conducive to beam loading. This causes an increasing energy spread due to head positrons experiencing larger fields than the tail (see Fig.\,\ref{fig:W_deltaW} later). This type of region where the bunch is accelerated but with increasing energy spread is termed as ``$\mathrm{I_{acc}}$". As protons have larger mass, their relativistic factors are essentially smaller in comparison with the ultra-relativistic positrons being further accelerated. Therefore, the protons and the wake phase keep shifting backwards with respect to the positrons. This seems creating a perfect opportunity for the positron bunch to get into the ``$\mathrm{D_{acc}}$" region ahead of the maximum accelerating field. The field here has a sharp and negative slope and is conceivable to rectify the energy spread. However, this region resides in the bubble head, where the plasma electrons stream across the axis and the plasma density is uneven. The radial fields thus vary transversely, which is detrimental to the beam emittance. 

Ref. \cite{yi2014positron} has still taken advantage of the ``$\mathrm{D_{acc}}$" region. But in there the single short proton driver is demanded dense enough so that an elliptical bubble nearly devoid of plasma electrons is formed behind the driver under strongly nonlinear wake excitation. That is, the perturbed plasma electrons move within a thin bubble sheath. In this way, a small radial space free from plasma electrons at the bubble head is maintained for the positron bunch. Nevertheless, it is not the case in our configuration. The multiple proton bunches will lose focusing from the plasma electrons and diverge significantly \cite{li2017multi}. This in turn leads to more plasma electrons flowing in and ruining the positron beam quality. Furthermore, with proton energy depletion and the bunch elongation after a long distance, it is increasingly tricky to maintain the plasma electrons closely along the bubble boundary. Fig.\,\ref{fig:ne_Ez_rz}c foresees the positron emittance growth from $z=84\,$m, when some plasma electrons get closer to the axis and the radial ``clean" space for positrons becomes smaller. The plasma electron trajectories crossing with that of positrons will dilute the positron bunch. However, by then the bunch has not reached the ``$\mathrm{D_{acc}}$" region (Fig.\,\ref{fig:ne_Ez_rz}g). Overall, there is a discrepancy between obtaining a small beam emittance and a small energy spread in our studied case. Basically the positron bunch needs to get to the bubble head to lower its energy spread gained before, nevertheless before that the nonuniform plasma electrons will destroy its emittance remarkably. As a result, either the acceleration is truncated before the beam emittance gets ruined or some measures must be taken from this point.

\subsection{Solution 1: plasma density decrease}
Recall that the ``$\mathrm{D_{acc}}$" region is conducive to the decrease of the energy spread due to the negative field slope. It follows that the ``$\mathrm{D_{dec}}$" region marked in Fig.\,\ref{fig:ne_Ez_rz}b should work in the same way. Although it is decelerating for positrons, it is plasma electrons free in a large radial space as it is located at the rear of the bubble ahead of the positron bunch. Thus, it is expected to be capable of conserving the beam emittance while reducing its energy spread. Since the bunch will deteriorate if it continuously slides to the ``$\mathrm{D_{dec}}$" region by means of phase dephasing, we propose to reduce the plasma density to $0.91n_{\mathrm 0}$ from the point when the beam emittance starts to degrade. In this way, the wake wavelength increases and the bubble shifts backwards (Fig.\,\ref{fig:ne_Ez_rz}d). It looks like the positron bunch directly jumps from the front of the old bubble to the tail of the bubble ahead (\textit{i.e.}, from Fig.\,\ref{fig:ne_Ez_rz}c to Fig.\,\ref{fig:ne_Ez_rz}d). 

It is noteworthy that while the plasma density change brings substantial phase shift in terms of the positrons, its effect on the resonance of proton bunches is acceptable. To be specific, the plasma wavelength $\lambda$ is inversely proportional to the square root of the plasma density $n$, therefore the slight density decrease $\delta n/n_{\mathrm 0}$ will cause the plasma wavelength to increase by $\delta\lambda/\lambda_0=0.5\delta n/n_{\mathrm 0}$. $\lambda_0$ is the initial plasma wavelength. It follows that with a relative density decrease of 9$\%$, the wake period increases by 4.5$\%$. This implies a negligible phase shift with respect to the first proton bunch considering the almost half wavelength it occupies. However, the phase shift accumulates from the first to the last bunch and it adds up to about 0.72 plasma wavelength relative to the last bunch. As initially the last bunch lags behind the maximum decelerating field (Fig.\,\ref{fig:ne_Ez_rz}b), the wake shift affects it insignificantly and most of protons still reside in the deceleration region. 

Note that there is already some dilution to the positron bunch in the large radius at the time of plasma density change. This is deliberate to keep the bunch staying longer in the initial acceleration stage with a large gradient and it is acceptable since the core beam (95$\%$) is still conserved. Fig.\,\ref{fig:W_deltaW} illustrates the evolution of the energy gain and energy spread of the core positron bunch in the initial acceleration stage (before $z=84\,$m) and the stage devoted to reducing the energy spread (after). Unfortunately, as seen in case 1, the decrease of the energy spread is essentially inefficient and even goes to the opposite way after a short distance when the bunch slips out of the ``$\mathrm{D_{dec}}$" region. Also it is with the cost of large energy gain loss. The reasons are threefold. First of all, the plasma density change although slight leads to less resonance of the proton bunches. Secondly, the proton bunches deform significantly after long term depletion. These result in less wake excitation and hence less sharp field slope compared to the initial stage (Fig.\,\ref{fig:ne_Ez_rz}b). Thirdly, the wake flattens more under the beam loading and the extension of the wake period (Fig.\,\ref{fig:ne_Ez_rz}h). Regardless, the positron beam emittance is still preserved. 

\subsection{Solution 2: presence of the electron beam load}
Since the plasma electrons are the main obstacle to the emittance preservation of the positrons, we introduce an extra electron bunch to the bubble tail ahead of the positron bunch (Fig.\,\ref{fig:ne_Ez_rz}e). The electron bunch has a large population of 2~$\times$~$10^{10}$. Its radial space charge force acts back on the bubble sheath electrons and delays them from returning to the $\xi$-axis. Therefore, the plasma electrons are expelled away from the positrons and the bubble shape is altered from that in Fig.\,\ref{fig:ne_Ez_rz}c to that in Fig.\,\ref{fig:ne_Ez_rz}e. Fig.\,\ref{fig:ne_Ez_rz}i shows that the loading of electrons changes the wake form as well and the positron bunch sees a flat or slightly increasing wakefield. Note here the electron bunch cannot be loaded to the decelerating field region as the energy loss will lead to its quick divergence under the only radial focusing of quadrupoles \cite{lotov2010simulation}. 

The new field slope at the beginning is not effective to reduce the energy spread, as it is still positive like before or almost zero. The negative slope ahead is expected to shift towards the positrons after a distance. The issue is by then the wake decreases due to more energy loss to both positrons and electrons and the field slope is less sharp. Fig.\,\ref{fig:W_deltaW} indicates that the loading of electrons (case 2) reduces the energy spread in a superior to the plasma density change case (case 1) but moderately efficient way. The good point is the positron energy increases continuously with preserved beam emittance until it gets further into the ``$\mathrm{D_{dec}}$" region. 

\subsection{Solution 3: combination}

The reason to the inefficiency of the second solution is when loading the electrons, the positrons are still far away from the ``$\mathrm{D_{acc}}$" region, whose slope becomes insignificant later when the positron bunch arrives. In consequence, we combine the aforementioned two approaches and propose to load extra electrons while reducing the plasma density to $0.95n_{\mathrm 0}$. By means of this, the positron bunch directly jumps to the ``$\mathrm{D_{acc}}$" region which has a large slope (Fig.\,\ref{fig:ne_Ez_rz}j) and meanwhile the presence of electrons removes the interference of plasma electrons with positrons (Fig.\,\ref{fig:ne_Ez_rz}f). More importantly, the required electron population is 1~$\times$~$10^{10}$ to maintain enough ``clean" space for positrons. This is twice smaller than in case 2 with only electron loading, where the electrons are further away from the positrons due to the requirement of staying in the accelerating field region and thus it requires a larger number of electrons to bend the bubble shape more. In addition, the plasma density decrease is less than in case 1 as it only needs to shift the ``$\mathrm{D_{acc}}$" region towards the positron bunch instead of the ``$\mathrm{D_{dec}}$" region further ahead. As a result, more protons are maintained within the decelerating region. 

\begin{figure}[!h]
\begin{minipage}[c]{17pc}
\subfloat{
   %\centering
   \includegraphics[width=12pc]{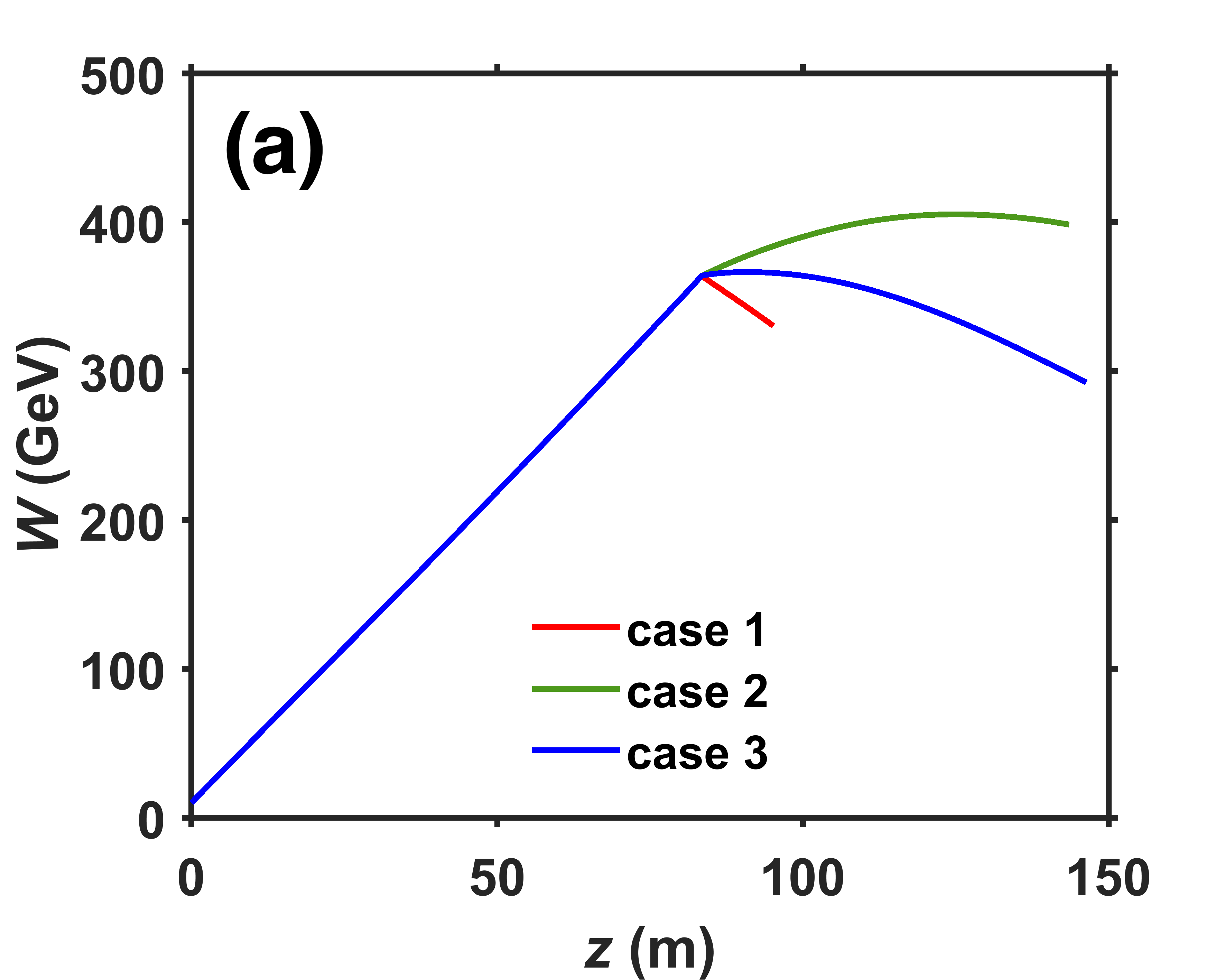}
  }
  \subfloat{
  % \centering
   \includegraphics[width=12pc]{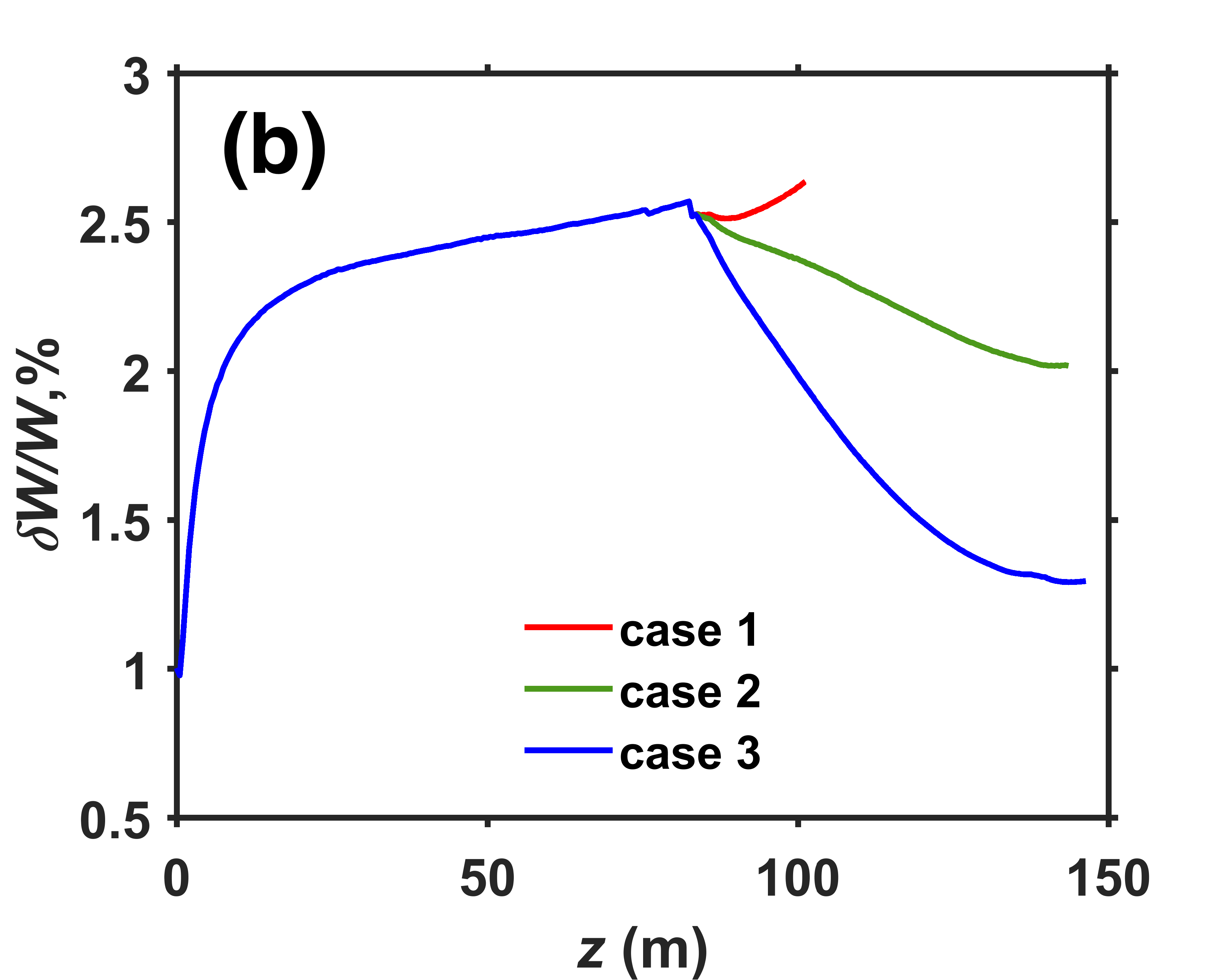}
  }
\end{minipage}\hfill
\begin{minipage}[c]{13pc}
\caption{Evolution of the positron energy gain (a) and energy spread (b) for the three proposed solutions starting working at $z=84\,{\mathrm m}$ to reduce the energy spread while preserving the beam emittance. Here it only shows the results of the core positron beam where the radial positions of positrons are within $0.1c/\omega_{\mathrm p}$ and the angular spread is smaller than 1e-6.}
 \label{fig:W_deltaW}
\end{minipage} 
\end{figure}

Figure\,\ref{fig:W_deltaW} (case 3) demonstrates the successful decrease of the energy spread to 1.3$\%$. Also it is substantially quicker than the other two cases. Over 95$\%$ positrons are kept at the initial normalized emittance (0.25 mm mrad). The net energy gain in the stage of reducing the energy spread is negative, which is inferior to case 2 as where the dephasing length seeing accelerating fields is longer. For all cases, the reduction of energy spread slows down when the positron bunch approaches the maximum decelerating field in the  ``$\mathrm{D_{dec}}$" region whose slope is trivial or almost zero. The end of this stage corresponds to the bunch getting out of the ``$\mathrm{D_{dec}}$" region, otherwise the energy spread will increase again.    

While accelerating the positrons, the electron bunch is accelerated from 10\,GeV to 150\,GeV as well but its energy spread is as large as 17$\%$ because it mostly staying in the ``$\mathrm{A_{acc}}$" region where the field is accelerating but the slope is not conducive. Its normalized emittance is well conserved in the whole process, as no plasma focusing acts on it. The further energy extraction by electrons suggests an increase of the overall energy efficiency for this particular configuration if this electron bunch is applicable. 

\section{Discussion}
We see the most efficient way to reduce the energy spread while preserving the beam emittance is to load an extra electron bunch and reduce the plasma density from the point when the plasma electrons start to interfere with the positrons. In this section, we further discuss the dependence of acceleration performance and beam quality preservation on the electron and plasma parameters. As to the electron beam, its charge and injection position need to satisfy the following conditions. Firstly, the charge is enough to expel plasma electrons away from the positron bunch. Secondly, the electron bunch resides in the acceleration region and better exactly ahead of the zero field (like the ones in Fig.\,\ref{fig:ne_Ez_rz}(i-j)) so that the energy extraction from protons is the least. 

\begin{figure}[!h]
 \begin{center}
  \subfloat{
   \centering
   \includegraphics[width=12pc]{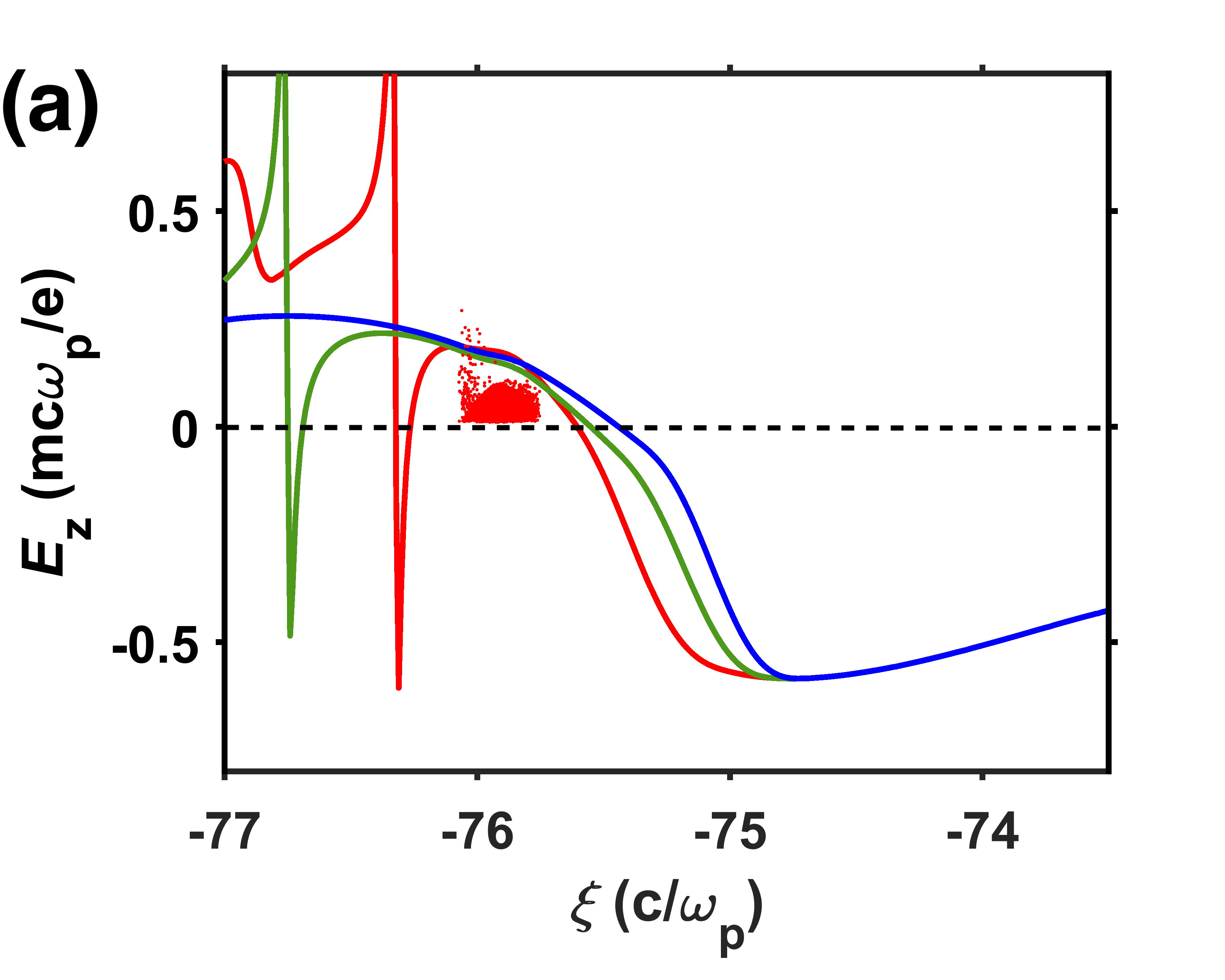}
  }
    \subfloat{
   \centering
   \includegraphics[width=12pc]{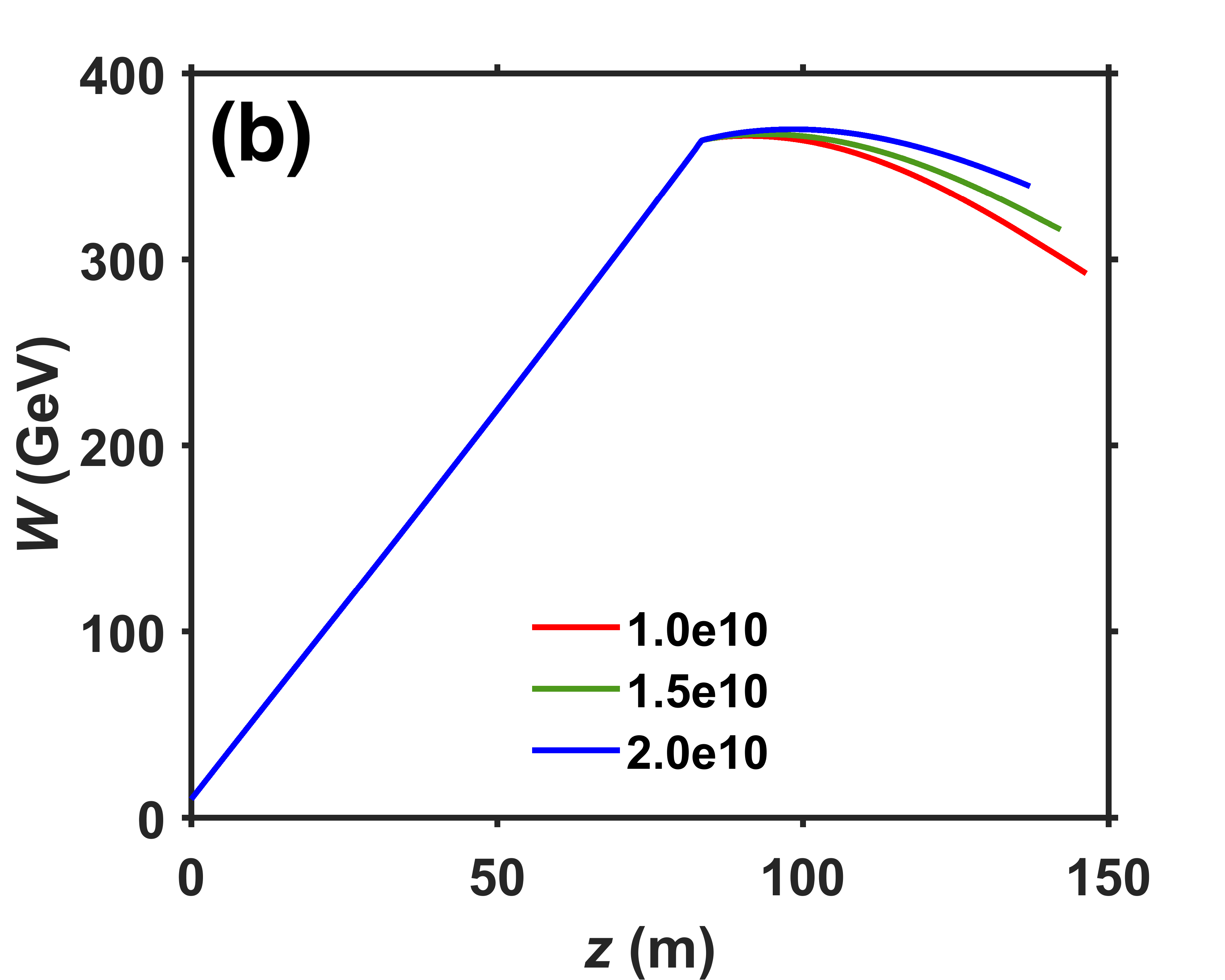}
  }
  \subfloat{
   \centering
   \includegraphics[width=12pc]{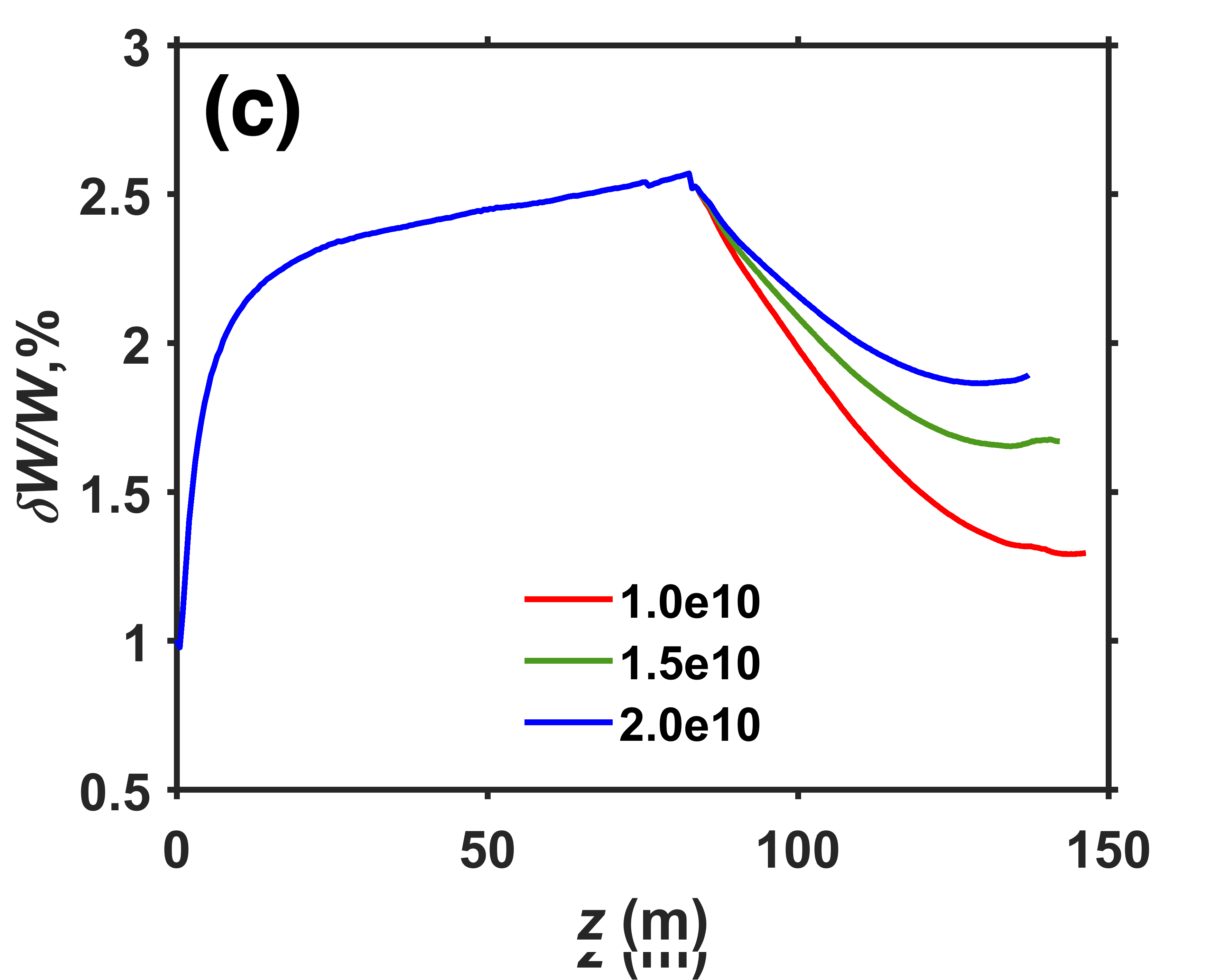}
  }
 \end{center}
 \caption{The on-axis longitudinal electric field seen by the positron bunch at $z=84\,$m when starting loading the extra electron bunch (a) and evolution of energy gain (b) and energy spread (c) for three cases of different electron charge loading under the same decreased plasma density of $0.95n_{\mathrm 0}$. The red points denote positrons. The electron bunch is not marked here.} 
  \label{diff_echarge}
\end{figure}

Given initially the same decreased plasma density of $0.95n_{\mathrm 0}$, Fig.\,\ref{diff_echarge} compares three cases under different populations of electrons. Apparently the electron bunch with larger charge resides further away from the positron bunch as more electrons overload the wake field to zero earlier. Therefore, it leaves a longer distance between the positron bunch and the zero field (Fig.\,\ref{diff_echarge}a), \textit{i.e.}, a longer dephasing length for acceleration before positrons getting into the decelerating field region. As a result, the net energy gain of positrons for the larger electron loading case is higher at the same distance (Fig.\,\ref{diff_echarge}b). Nevertheless, the wake field slope seen by the positron bunch is smaller and thus the energy spread drops slower (Fig.\,\ref{diff_echarge}c).

\begin{figure}[!t]
 \begin{center}
  \subfloat{
   \centering
   \includegraphics[width=12pc]{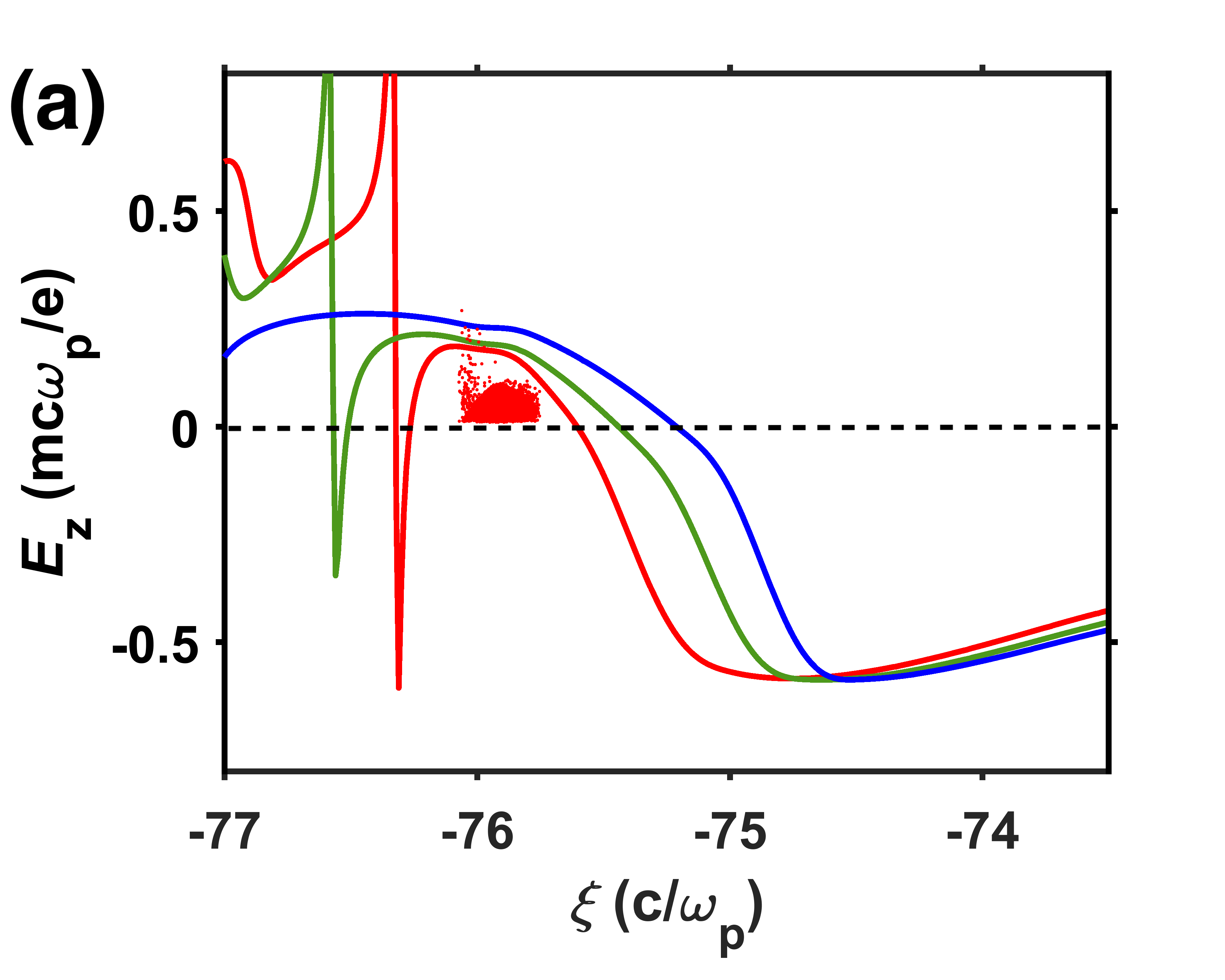}
  }
    \subfloat{
   \centering
   \includegraphics[width=12pc]{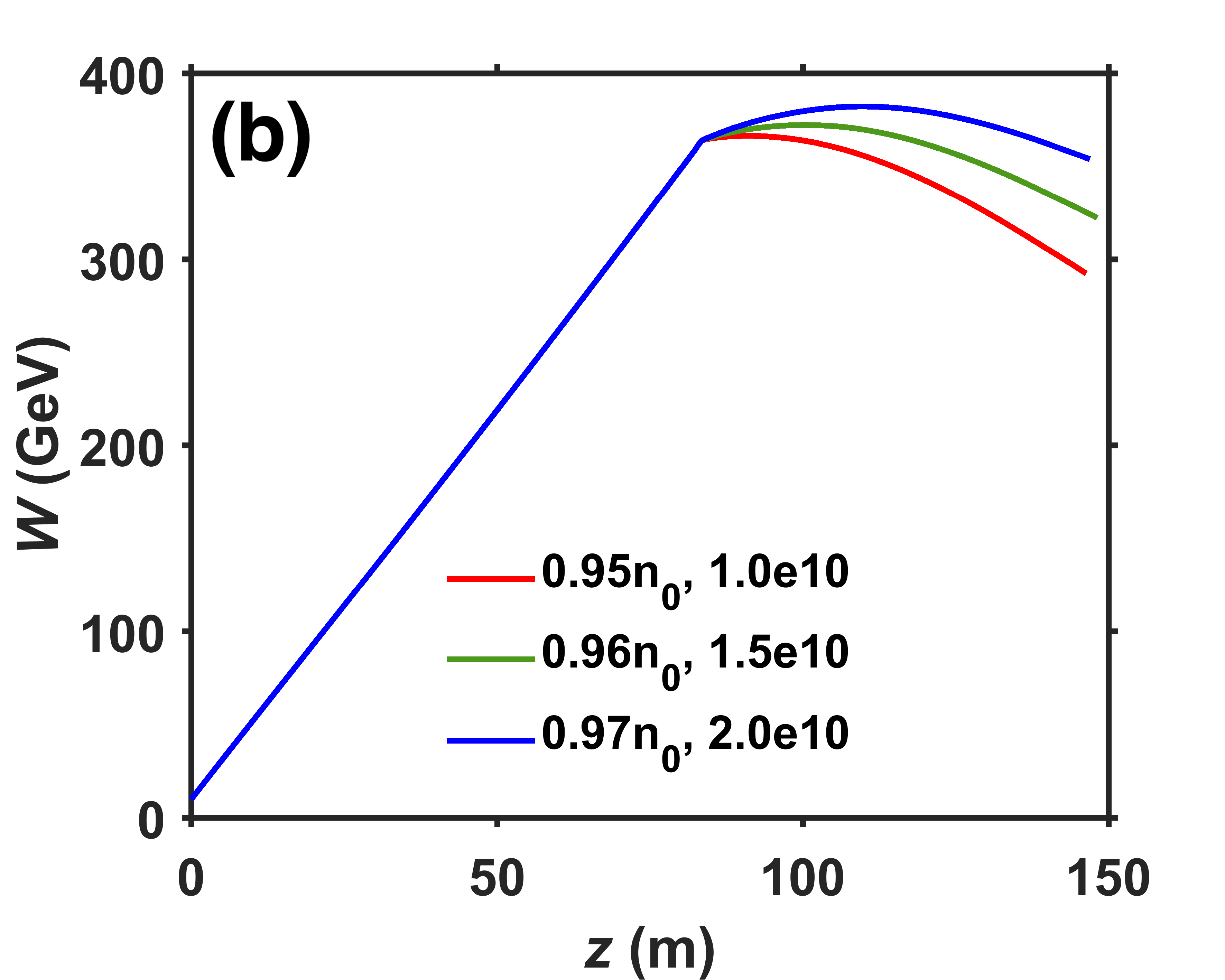}
  }
  \subfloat{
   \centering
   \includegraphics[width=12pc]{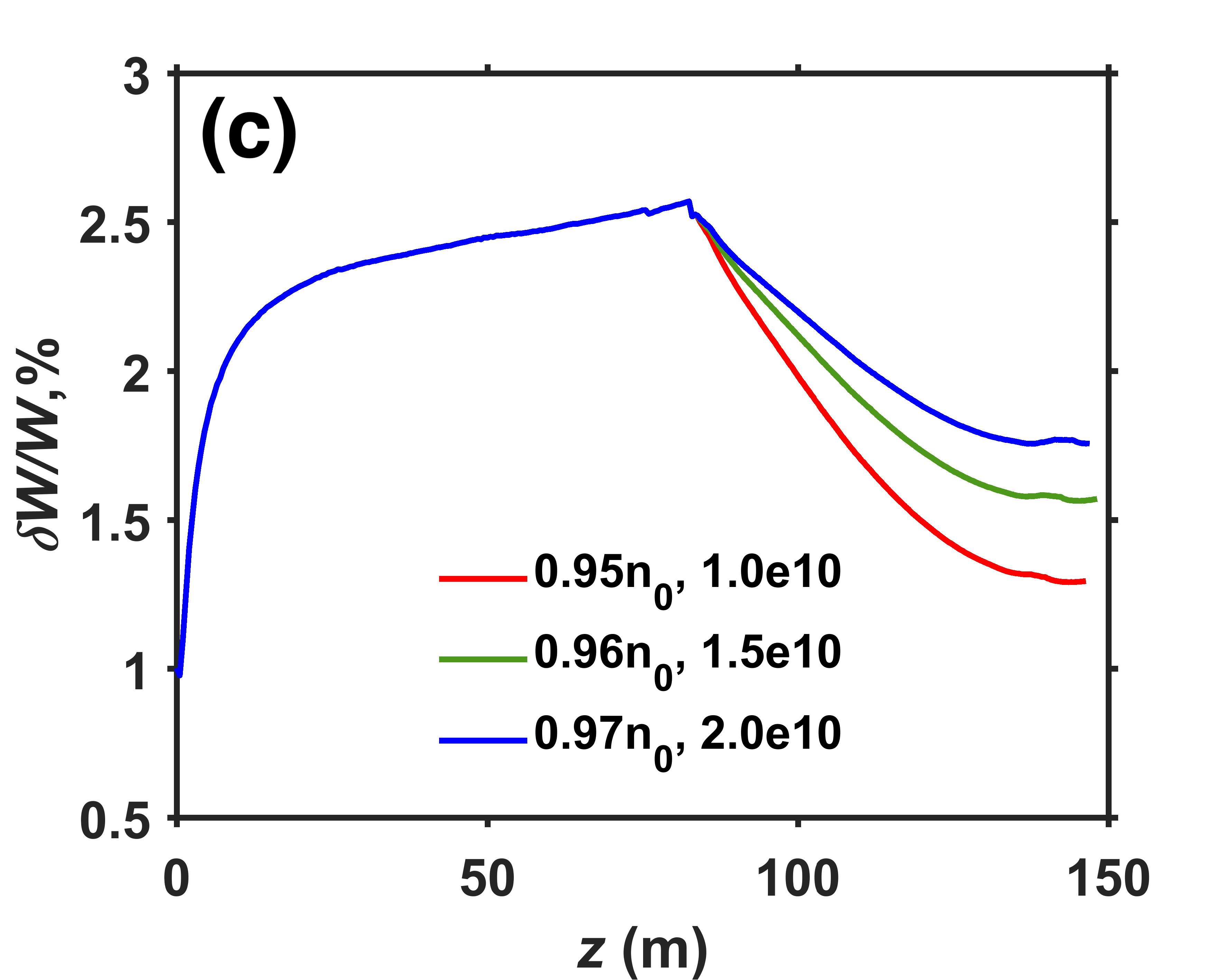}
  }
 \end{center}
 \caption{The on-axis longitudinal electric field seen by the positron bunch at $z=84\,$m when starting loading the extra electron bunch (a) and evolution of energy gain (a) and energy spread (b) for three cases of different electron charge loading accompanied by different decreased plasma densities. The red points denote positrons. The electron bunch is not marked here.} 
  \label{diff_echarge_np}
\end{figure}

It has been demonstrated that a dropped plasma density of $0.95n_{\mathrm 0}$ shifts the ``$\mathrm{D_{acc}}$" region well to the positrons.  Although with a larger plasma density the effect of phase shift on multiple proton bunches will be less, because the electron bunch is further away from the positron bunch, it requires a larger electron loading so that the space charge force is enough to repel the plasma electrons more. Fig.\,\ref{diff_echarge_np} indicates the cases where the larger dropped plasma density comes with a larger electron load. Similar to the case in Fig.\,\ref{diff_echarge}, the energy spread decreases slower with a large electron load but the net energy gain increases. Also when comparing Fig.\,\ref{diff_echarge} and Fig.\,\ref{diff_echarge_np}, it is apparent that with the same electron loading, the wakefield and energy gain are promoted under a larger dropped plasma density. This is because more protons contribute to the wake excitation under less phase shift. 

In Section 2, we see the net energy gain in the third solution (over the stage of reducing energy spread) is negative because of positrons later sliding into the ``$\mathrm{D_{dec}}$"  region with wake phase shift. Then it is natural to think further increasing the plasma density so that the wake wavelength decreases and the ``$\mathrm{D_{acc}}$" region is shifted back to the positron bunch. Meanwhile, with wake wavelength decrease, the extra electron bunch is expected to move towards a smaller acceleration gradient and extract less energy from the plasma wave. 

\begin{figure}[!t]
\begin{minipage}[c]{22pc}
  \subfloat{
   \centering
   \includegraphics[width=22pc]{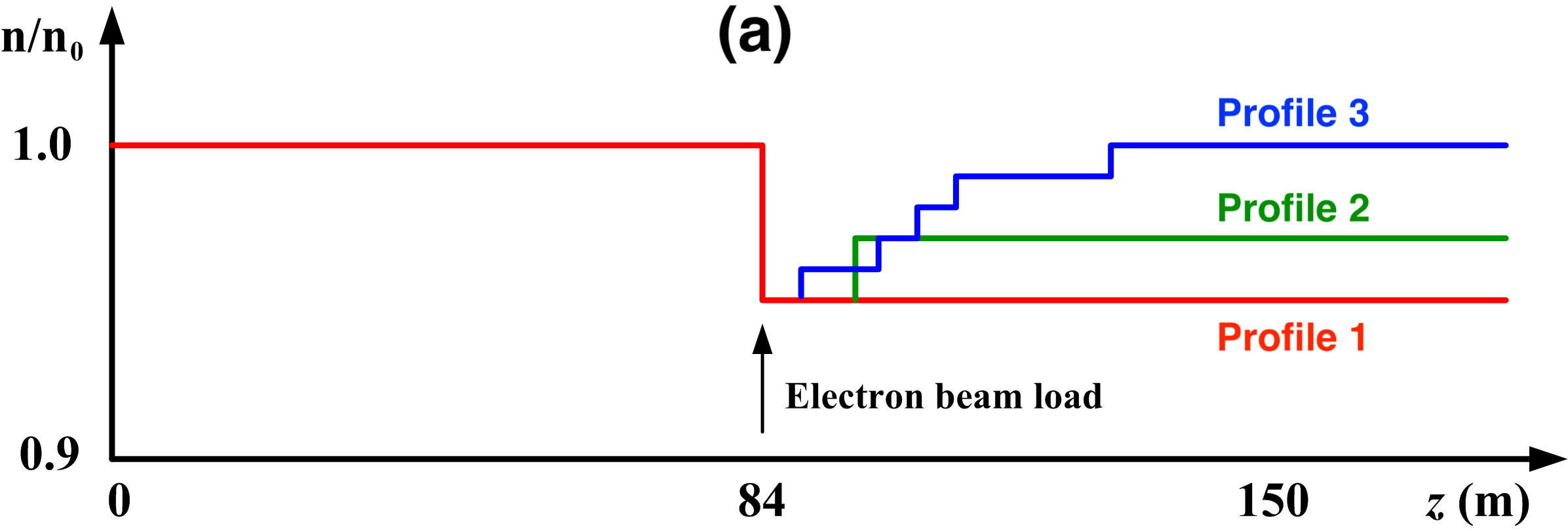}
  }
\newline
  \subfloat{
   \centering
   \includegraphics[width=11pc]{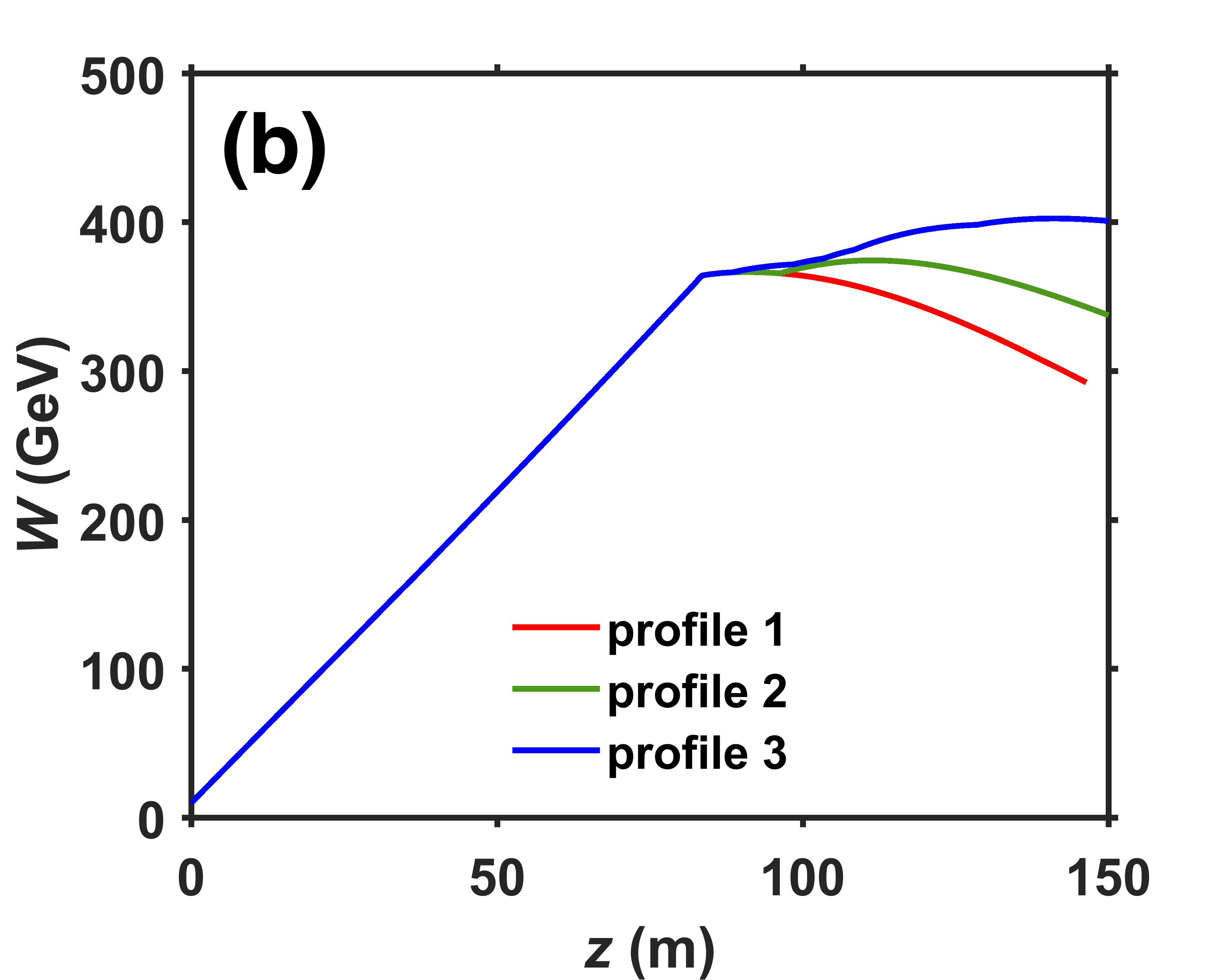}
  }
  \subfloat{
   \centering
   \includegraphics[width=11pc]{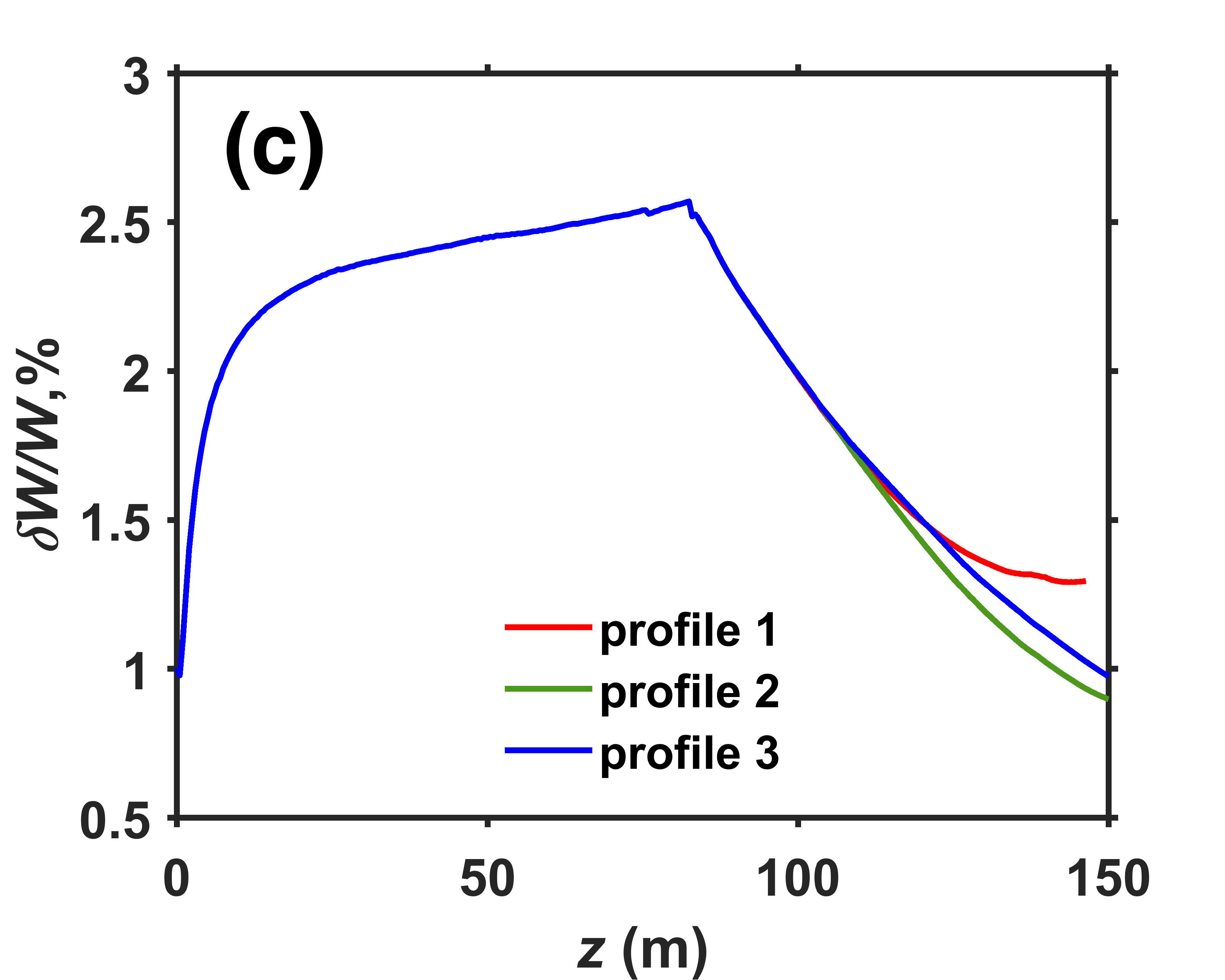}
  }
\end{minipage}\hfill
\begin{minipage}[c]{14pc}
 \caption{\label{label}Three different types of longitudinal plasma profiles (a) and the corresponding evolution of energy gain (b) and energy spread (c) of positrons being accelerated in such plasma channels. The identical colour represents the same case. The red one corresponds to the third solution presented in section 2.}
 \label{diff_profiles}
\end{minipage}
\end{figure}

Figure\,\ref{diff_profiles} compares the cases with different longitudinal plasma profiles adopted along with the same electron beam loading (1~$\times$~$10^{10}$). The red coloured case represents the third solution discussed in Section 2. With a slight plasma density increase from $0.95n_{\mathrm 0}$ to $0.96n_{\mathrm 0}$ (\textit{i.e.}, the green line in Fig.\,\ref{diff_profiles}a), both the net energy gain and the decrease rate of energy spread are promoted (Fig.\,\ref{diff_profiles}(b-c)). Still, for a longer acceleration distance, the positrons will inevitably get into the ``$\mathrm{D_{dec}}$"  region again. Therefore, a more complicated plasma profile (\textit{i.e.}, the blue line in Fig.\,\ref{diff_profiles}a) is proposed where the plasma density increase occurs when or after the positrons slip into the ``$\mathrm{D_{dec}}$"  region. In this case the positron energy increases continuously to 400\,GeV while the energy spread decreases to the initial level of 1$\%$ or even less with a longer distance at almost the same rate. Here the plasma with a stepped density can be realized by segmented plasma in practice. In addition, it allows a substantial response distance to change the plasma density, as the wake phase changes insignificantly in terms of the acceleration distance. For instance, the wake only dephases by around 8.4e-3$c/\omega_{\mathrm p}$ if passing by 1\,m longer plasma. Further simulations prove that the density change is not necessarily sharp and a gradual change even helps on the acceleration performance.

\section{Conclusions}

In this paper, we firstly analyse the issues in preserving the quality of the positron bunch while accelerating it to energy frontier in the multiple proton bunch driven hollow plasma wakefield accelerator. There is a discrepancy between keeping a small normalized emittance and a small energy spread. This results from the conflict that the plasma electrons providing focusing to the multiple proton bunches dilute the positron bunch. Then we propose and compare three different solutions, and prove that the most efficient way is to load extra electrons and meanwhile decrease the plasma density from the point when the positron emittance starts to degrade. Simulations show that the positron bunch can be accelerate to 400\,GeV with the energy spread as low as 1$\%$ when adopting a slightly more sophisticated plasma profile to keep it more in the ``$\mathrm{D_{acc}}$"  region. Its normalized emittance is well preserved at the initial value. The electron beam load is crucial and it needs to accommodate well with the dropped plasma density to ameliorate the acceleration performance. For a larger decreased plasma density, the electron beam loading must be larger. Although the corresponding energy spread decrease slows down, the net energy gain increases. This work expands the concept of positron acceleration driven by protons and the obtainable high quality and high energy positrons are promising candidates for the future energy frontier lepton colliders.

\section*{Acknowledgements}
The authors appreciate the financial support from the President's Doctoral Scholarship Award of the University of Manchester, the Cockcroft Institute core grant and STFC. The authors greatly thank the computing resources provided by STFC Scientific Computing Department's SCARF cluster and also by the CSF cluster at the University of Manchester. 
%\iffalse  % only for "biblatex"
%	\newpage
%	\printbibliography
%
%% "biblatex" is not used, go the "manual" way
%\else

\section*{References}
\bibliography{myiopart-num}

\end{document}